\documentclass[11pt]{article}

\usepackage{geometry}
\usepackage[utf8]{inputenc} 
\usepackage[T1]{fontenc}    
\usepackage{hyperref}       
\usepackage{url}            
\usepackage{booktabs}       
\usepackage{amsfonts}       
\usepackage{nicefrac}       
\usepackage{microtype}      
\usepackage{xcolor}         
\usepackage{graphicx}
\usepackage{algorithmic}
\usepackage{algorithm2e}
\usepackage{amsmath}
\RestyleAlgo{ruled}
\SetKwComment{Comment}{/* }{ */}

\usepackage{multirow}

\title{Aggregation of Constrained Crowd Opinions for Urban Planning}

\author{%
  Akanksha Das, Jyoti Patel and Malay Bhattacharyya \\
  Indian Statistical Institute \\
  Kolkata -- 700108 \\
  India \\
  E-mail: akankshadas2020@gmail.com, jyotiinfinite72@gmail.com, malaybhattacharyya@isical.ac.in \\
}

\begin{document}

\maketitle

\begin{abstract}
Collective decision making is often a customary action taken in government crowdsourcing. Through ensemble of opinions (popularly known as judgment analysis), governments can satisfy majority of the people who provided opinions. This has various real-world applications like urban planning or participatory budgeting that require setting up {\em facilities} based on the opinions of citizens. Recently, there is an emerging interest in performing judgment analysis on opinions that are constrained. We consider a new dimension of this problem that accommodate background constraints in the problem of judgment analysis, which ensures the collection of more responsible opinions. The background constraints refer to the restrictions (with respect to the existing infrastructure) to be taken care of while performing the consensus of opinions. In this paper, we address the said kind of problems with efficient unsupervised approaches of learning suitably modified to cater to the constraints of urban planning. We demonstrate the effectiveness of this approach in various scenarios where the opinions are taken for setting up ATM counters and sewage lines. Our main contributions encompass a novel approach of collecting data for smart city planning (in the presence of constraints), development of methods for opinion aggregation in various formats. As a whole, we present a new dimension of judgment analysis by adding background constraints to the problem.
\end{abstract}

\section{Introduction}
Crowdsourcing has shown an immense promise in effectively employing the power of human (termed as the crowd, be a collection of experts, non-experts, or their mixture) to solve a variety of real-world problems at scale during the last couple of decades \cite{howe2006rise,kittur2013future,lenart2023understanding}. Judgment analysis is an important mechanism in crowdsourcing that involves collecting opinions from the crowd, followed by yielding a consensus of the opinions  to address a problem \cite{chatterjee2019review}. This mechanism is itself a challenging computational problem to solve, given its newer dimensions added in the area of crowdsourcing, as opposed to the traditional judgment analysis introduced much earlier \cite{dawid1979maximum}.

Unlike traditional judgment analysis, where we need domain experts, systems that require to perform judgment analysis on crowdsourced data employs common people. Opinions in this form put forward a new set of challenges because of the sparsity and multi-dimensionality of opinions. There are earlier attempts to develop a generalized approach of judgment analysis toward dealing with multi-dimensional opinions received from the crowd \cite{chatterjee2019review}. However, obtaining the consensus of multi-dimensional opinions is still challenging because of the latent associations between the different dimensions of an opinion. Generally, these associations are verified by setting up constraints between the opinions \cite{chatterjee2017constrained,chatterjee2017smart}, thereby forming the constrained judgment analysis problems. But unfortunately, they can never ensure the collection of reliable opinions from the crowd. Moreover, a simple consensus of opinions does not guarantee providing a valid option as the final judgment. In this work, we address this issue by developing reliable constrained judgment analysis by adding background constraints to the problem. This approach helps to obtain a better consensus of multi-dimensional opinions with hidden constraints taken into consideration.

To better understand how the general constraints, which are traditionally used in constrained judgment analysis, are treated separately from the newly introduced constraints. We formally define these two types of constraints as follows.

\begin{itemize}
    \item \textbf{Main Constraints:} The main constraints are basically restrictions designated between only the opinions to be collected from the crowd.
    \item \textbf{Background Constraints:} The background constraints are basically restrictions designated between the opinions and the background information.
\end{itemize}

We have applied and experimented this new version of constrained judgment analysis with the help of two different case studies. We have demonstrated the utility of this model for strategic planning of smart cities in general. Additionally, we have put forth a paradigm that can handle complex viewpoints while upholding limitations. The novelty of the current paper lies in the following things.

\begin{enumerate}
\item We introduce a novel problem related to crowd judgment analysis where multiple types of constraints are taken into account.
\item We propose a novel opinion aggregation approach that can deal with different forms of opinions (points or lines).
\item The efficacy of the proposed method is shown by applying it on both point and line based opinion aggregation.
\end{enumerate}

As a whole, the current work bears novelty in both methodology and applications.

\section{Basic Terminologies}
Let us introduce the basic terminologies that will be used throughout the paper. We involve four major terms, namely question, annotator, opinion, and gold judgment.

A \emph{question} is simply a labeling task. E.g., it can be the annotation of an image or a text (like tweets, online posts, etc.). In the current work, the questions are related to locations of objects.

An \emph{annotator} is essentially the crowd worker who provide the opinion (annotates a label) over the given question. Annotation can be thought of as a function that returns one of the available options according to some procedure. An annotator may be paid for answering the questions or can be volunteers. In the current work, the annotators are volunteers.

An \emph{opinion} is the judgment given by an annotator for a particular question. Generally, the opinion should be chosen from an opinion set. However, in the current work, the opinions are chosen from a bounded but continuous space in $\mathbb{R}^2$. The opinions are represented as coordinates, the question being related to a point or a line.

The \emph{gold judgment} is the original true label of the question. The questions are distributed among the annotators soliciting their individual opinions. Then the aggregation is done to predict the gold label. The deviation of the predicted label from the gold label can be easily computed. In the current work, the gold judgments are not defined. Therefore, the idea of obtaining a gold judgment is exploratory by nature.

\section{Motivation}
Our major motivation in this work is collecting opinions from the crowd workers in a way such that random inputs are avoided as much as possible. The purpose of collecting opinions is setting up facilities in a smart city. The smart city planning considers the inclusion of the said facilities in many different forms like the establishment of ATM counters, sewage lines, etc.) based on the opinions received from the crowd. As the opinions might be points, lines, polygons, etc., we need a flexible approach to deal with the aggregation problem. Moreover, the constraints are learned from the practical setting of the problem.

\section{Related Work}
A good summary of different judgment analysis problems can be found in a review by Chatterjee et al. \cite{chatterjee2019review}. The problem of aggregating constrained judgments was first introduced in a legal setting \cite{dietrich2010problem}. Dietrich et al. added explicit constraints to the opinions bringing the rationality conditions of consistency in the picture. The variations of this somewhat makes the notions of rationality flexible in the exogenous constraints. This paper presents some interesting theoretical results on constrained judgment aggregation. However, this approach was not generalized, and was unable to consider relationships between and reliability of opinions.

In many tasks, like protein structure prediction the relationship among various options are taken into consideration \cite{peng2013}. Peng et al. attempted to bring constraints into the crowd judgment judgment analysis in an associative way. However, it assumes that most of the workers are reliable and annotators can be easily grouped based on their opinion patterns. Though the formation of similarity matrices between annotators is meaningful for protein structure prediction, it cannot be used in a generic way for other problems. Moreover, these formulations are easy to be addressed with majority voting (to be specific kernelized majority voting). But for other problems, where opinions are in more complex forms (say coordinates), majority voting cannot ensure the satisfiability of the specified constraint.

Chatterjee et al. introduced the concept of a more complex form of opinions and added constraints between the opinions taken \cite{chatterjee2017smart}. This was basically done for smart city planning, however consequently, many other possible applications were also shown \cite{chatterjee2017constrained}. This was about considering a consensus of points where the input points must satisfy the constraints. There were also the same constraints applicable to the final consensus obtained. However, this approach took a filtering strategy to remove the opinions that dissatisfy the constraints. First of all, this reduces the number of opinions. Secondly, it does not ensure a reliability of opinions by restricting the workers. In this way, constrained opinions are managed entirely in an ad hoc basis for a particular application. In this work, where the opinions are multiple ordinates of a location represented with real values, a Bayesian binning approach is adopted for resolving validity problems of opinions and then a probabilistic graphical model is applied. However this work, on smart city planning, does not provide a general framework for judgment analysis on multi-dimensional opinions where many other latent relations are possible. In recent years, constrained judgment analysis problems have been attempted using metaheristic approaches \cite{chatterjee2020multi,chatterjee2022topsis}. However, no new dimension of the problem has been added.

Unfortunately, there is no existing work that makes use of dependency between multi-dimensional opinions and background details to develop a generic opinion elicitation model, which is one of the major goals of this work. Our work provides a generalized framework to deal with constrained opinions toward judgment analysis aiming various real-life applications. Unlike the existing approaches of judgment analysis, we also aim to collect opinions in a way such that annotators are made responsible toward their opinions.

There is a close connection of our work, which considers opinions in different forms, with the problem of clustering of objects \cite{duda2006pattern}. As we consider points and lines as opinions, clustering of lines becomes a challenging problem to deal with. There are limited attempts in the literature to form clusters of lines. Peretz was the first to bring the idea of extending centroid-based clustering to lines \cite{peretz2011thesis}. Leter Marom et al. have extended this to more interesting applications \cite{marom2019a,marom2019b}. Our problems are indeed novel, however, the proposed methods are inspired by these works.

\section{Empirical Analysis on Urban Planning}
Our primary contribution is effectively incorporating background constraints in constraint judgement analysis. The constraints (main and background) are specific to the type of facilities to be set up and the format in which the opinions (points or lines) are collected from the crowd workers. We exhibit how the background infrastructure can included in the data collection process to reduce the amount of erroneous data and how increasing the amount of background information automatically incorporates certain background constraints that are difficult to verify after the data is collected (see Case Study 1). In addition to that, instead of removing the data-points that do not satisfy the constraints, we adjust the data points slightly wherever possible such that it retains the essence of the original opinion but satisfies the constraints (see Case Study 2). Traditionally, judgement analysis has been done by assuming that the opinions collected from the crowd workers are in the form of points but in certain applications (see Case Study 2), it is important to denote the opinions as lines instead of points. We suggest a model inspired from k-median of lines with Hausdorff distance \cite{marom2019a, marom2019b}, and discuss its benefits over the distance measure used by Marom et al. \cite{marom2019a, marom2019b}.

It is important to note that the constraints that we wish to consider are purely situation specific, we exhibit our contributions through the following two case studies of establishment of ATMs and Sewage Planning deals with point opinions and line opinions respectively (see Table~\ref{Table:Summary}). In Case Study 1, 38 crowd workers based in an office in Chandni Chawk, Kolkata suggest 6 ATM locations each, 3 of them without the knowledge of existing ATMs of their preferred banks in the area and the rest with the knowledge of existing ATMs of their preferred banks in the area. We do a comparative analysis on the two sets of opinions collected. In Case Study 2, 50 respondents from all over Kolkata propose 3 unique sewage lines each. In Case Study 2, 50 respondents from all over Kolkata propose 3 unique sewage lines each. We study how faulty opinions may be adjusted and/or retained  while still yielding a constraint satisfying consensus. In each of the following case studies, data has been collected from a diverse group of people with varying levels of education. The statistical details of the crowd workers are summarised in Table~\ref{Table:Statistical}.   

\begin{table}[htp]
    \centering
    \begin{tabular}{|c|c|c|c|}
        \hline
        \textbf{Case Study} & \textbf{Opinion} & \textbf{Facility} &\textbf{\# Respondents} \\
        \hline
        1 & Point & ATM & 37 \\
        2 & Line & Sewage line & 50 \\\hline
    \end{tabular}
    \caption{Summary of the problems and data collected under the Case Study 1 and 2.}
    \label{Table:Summary}
\end{table}

\begin{table}[htp]
    \centering
    \begin{tabular}{|c|c|c|p{0.7in}|}
        \hline
        \textbf{Case} & \textbf{Gender} &\textbf{\# Graduate} & \textbf{Location}\\
        \textbf{Study} & \textbf{Ratio (F:M)} & \textbf{Respondents} &  \\
        \hline
        1 & 1:12 & 31 ($\approx 82\%$) & 10, Madan Street, Chandni Chawk \\
        2 & 1:4 & 39 ($78\%$) & Greater Kolkata \\\hline
    \end{tabular}
    \caption{Statistical and background details of the crowd workers employed in the two different case studies.}
    \label{Table:Statistical}
\end{table}

\section{Case Study 1: Establishment of ATMs}
Our principal aim is to establish 3 ATMs based on the opinions given by the crowd worker and each crowd-worker is allowed 3 opinions. This problem has been explored in constraint judgement analysis with certain main constraints, which are constraints on the opinions only \cite{chatterjee2017smart}. We introduce certain background constraints in the problem and analyse the effect of background details, included while collecting the data. We collect the opinions in the form of coordinates (pair of latitude and longitude) corresponding to the location of the ATMs, desired by the crowd workers on Google Maps both with and with out providing the information of the location of existing ATMs to the crowd workers. This aids us in comparing the data obtained in the above scenarios by the same crowd worker and observations from the comparative analysis provide us with insight on how to significantly reduce the percentage of data that violates the constraints imposed, hence it would allow the retention of more opinions (that would otherwise be removed) and subsequently lead to a better consensus. We specify the main and background constraints below.

\textbf{Main Constraints \cite{chatterjee2017smart}:}
\begin{itemize}
    \item \textbf{Boundary:} The proposed ATM should lie in the Chandni Chawk Area. 
    \item \textbf{Intra-Separability:} The three ATMs proposed by a crowd worker should be at least 100m apart from each other to ensure the uniqueness of the opinions.   
\end{itemize}

\textbf{Background Constraints:}
\begin{itemize}
    \item \textbf{Inter-Separability :} Every opinion of a crowd worker must be at at least some distance apart from the existing ATMs of the same type unless the existing ATM is not in an open space. This ensures that the new ATMs do not overlap with the existing ones.  
    \item \textbf{Location :} The location of ATMs should be in an open space, preferably on the roadside.
    \item \textbf{Preferential Treatment :} Preference should be given to the opinions that do not have existing ATMs of the same type.
\end{itemize}

\subsection{Data Collection}
The data was collected from 37 workers of an office in 10, Madan Street, Chandni Chowk, Kolkata. The respondents belong to different age groups and have varying educational qualifications. Each crowd worker suggests 3 locations of ATMs of the desired bank on the Google Map of Madan Street, Chandni Chawk and 3 locations of ATMs of the above type after the existing ATMs of that type are marked on the map. Other than that, we have also discussed with the crowd workers why do they think the existing ATMs are not sufficient and the challenges they face because of it. Their views are presented in the \textbf{Discussion section}. We have decided on the aforementioned constraints and model based on our conversations with the crowd workers. We have obtained 2 datasets with 37 entries each namely, ATM1 and ATM2 where ATM1 deals with the data collected without the associated background details while ATM2 deals with the data collected with background details.

\subsection{Handling the Background Constraints}
\textbf{Constraint Analysis in ATM1:} Data in ATM1 already includes background details like location of roads, Chandni Chawk area demarcation hence all its opinions automatically follow the  Boundary and Location constraints. In addition to that, none of the crowd workers have repeated their opinion. This may have been the result of our dataset having $82\%$ graduate respondents. As some of our crowd workers were not aware of the explicit location of ATMs, we have some violations ($\approx 5\%$) of the Inter-separability constraint. 

\textbf{Constraint Analysis in ATM2:} Data in ATM2 includes background details like location of roads, Chandni Chawk area demarcation hence all its opinions automatically follow the  Boundary and Location constraints. In addition to that, none of the crowd workers have repeated their opinion. This may have been the result of our dataset having $82\%$ graduate respondents. Contrary to ATM1, all our crowd workers are aware of the explicit location of ATMs in the vicinity hence we have no violations of Inter-separability constraint.

Note that the Preferential Treatment constraint has been dealt with in the model of Judgement Analysis. 

\begin{table}[htp]
    \centering
    \begin{tabular}{|c|c|c|c|}
    \hline
    \textbf{Dataset} & \textbf{\# Proposed} &\textbf{\# Opinions Violating} & \textbf{Error}\\
    & \textbf{ATMs} & \textbf{Inter-separability} &  \\
    \hline
    ATM1 & 111 & 6 & $\approx 5\%$ \\
    ATM2 & 111 & 0 & $0\%$ \\\hline
    \end{tabular}
    \caption{Analysis of constraints pertaining to ATM1 and ATM2.}
    \label{Table 3}
\end{table}

\begin{table}[htp]
    \centering
    \begin{tabular}{|c|c|c|c|}
        \hline
        \textbf{Associated} & \multicolumn{2}{|c|}{\textbf{\# Existing ATMs}} & \textbf{\# Proposed} \\\cline{2-3}
        \textbf{Bank} & \textbf{within 300m} & \textbf{within 750m} & \textbf{ATMs} \\
        \hline
        SBI & $1$ & $1$ & $51$ \\
        AXIS & $2$ & $>5$ & $24$ \\
        ICICI & $1$ & $2$ & $21$ \\
        BOB & $0$ & $1$ & $6$ \\
        HDFC & $1$ & $>5$ & $6$ \\
        IDBI & $2$ & $2$ & $3$ \\\hline
    \end{tabular}
    \caption{Number of existing and proposed ATMs of different banks near 10, Madan Street, Chandni Chawk.}
    \label{Table 4}
\end{table}

\subsection{Model of Judgment Analysis}
After removing the erroneous data (if any as in ATM2), we have to decide the the banks of the ATMs that have to be constructed. From Table 4, we can see that the proposed ATMs of different banks are in the ratio 17:8:7:3:3:1, hence the ATMs should be in the same ratio i.e 2 ATMs of SBI and 1 ATM of AXIS bank but as we can see ICICI bank has almost the same number of proposed ATMs as Axis Bank. Hence, here we employ the constraint of Preferential Treatment. AXIS bank has sufficient number of existing ATMs in the vicinity hence we chose ICICI bank as both SBI and ICICI have less than 3 ATMs within 750m. Hence, we construct 2 ATMs of SBI and 1 ATM of ICICI bank. Now, we apply the constrained judgement analysis algorithm propsoed by Chatterjee et a.. \cite{chatterjee2017smart} separately on the 51 ATMs of SBI to get 2 ATM locations and 21 ATMs of ICICI to get 1 ATM location. This is our final consensus.

\subsection{Results and Observations}
Though the same crowd worker presents two sets of opinions, they are almost always significantly different from each other. In the first scenario, the crowd worker almost always thinks of his personal interests and marks the exact locations where he/she is most likely to withdraw money from. On the other hand, knowing the locations of existing ATMs, makes the crowd worker something of a town planner himself who tries to base the location of the ATM that is also aligned to the community need along with his/her personal requirements. This leads to a more responsible set of opinions and hence a better consensus. This also reduces the amount of random opinions as crowd workers with no specific need for ATMs can base their opinions on the location of existing ATMs. Hence, a detailed background while collecting data is key to the right consensus.

\section{Case Study 2: Sewage Planning}
In this problem, we consider the opinions in the form of line segments, precisely on a map.

\subsection{Problem Statement}
We demarcate the area given for sewage planning and construction as a convex region. We consider the Euclidean distance measure between any two points (in $\mathbb{R}^2$) to symbolize their closeness. For simplicity, we denote a sewage line in terms of the coordinates of its end points. Let there be $k$ existing background sewage lines provided beforehand inside the given region. To set up additional sewage lines, we seek opinions from the crowd workers, where each crowd worker suggests $k^*$ number of new sewage lines. We have to find their consensus satisfying some constraints such that they deem fit in the model as reliable opinions.

\begin{figure}[htp]
    \centering
    \includegraphics[height=2.4in]{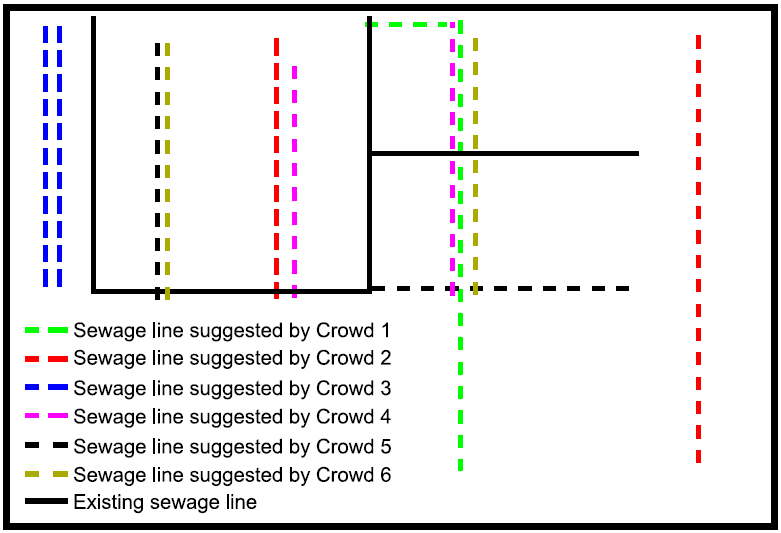}
    \caption{An example scenario of judgment analysis with background constraints is shown. The outside boundary denotes the area under consideration for sewage planning and construction. The problem aims to set up new sewage lines in a city by taking suggestions (opinions) from the crowd. We consider that each crowd provides two opinions in this regard. Judgments are to be derived by computing consensus of the suggested sewage lines (shown as dashed lines) opined by the crowd. Background constraints are related to the sewage lines already in existence (shown as solid lines). In this example, the background constraint is: (1) Each suggested sewage line must have at least one intersection point with the existing sewage lines, and the additional general constraints are: (2) Each suggested sewage line must have a maximum length, and (3) A pair of suggested sewage lines must have a minimum distance between themselves. Note that the opinions given by the crowd 4, 5 and 6 satisfy all the constraints and are valid. However, the opinions of crowd 1 does not satisfy the constraints (2) and (3), the opinions of crowd 2 does not satisfy the constraints (1) and (2), and the opinions of crowd 3 does not satisfy the constraints (1) and (3). Hence, the final judgment is obtained from the consensus of valid opinions provided by the crowd 4, 5 and 6 only.}
    \label{Fig:Example}
\end{figure}

To exemplify the problem, let us consider that we have a smart city under development (as shown in Fig.~\ref{Fig:Example}), and we want to construct sewage lines based on opinions from the crowd, with citizens participating as annotators. The boundary area represents the city being designed. The annotators provide their opinions on where the sewage lines should be constructed, represented by dashed lines. These opinions form the suggested paths for the sewage lines.

There are certain constraints in place for constructing the sewage lines. Firstly, each sewage line should have a maximum length to ensure efficient and manageable infrastructure. Additionally, there is a requirement for a minimum distance between two sewage lines to avoid congestion and optimize the layout of the city. However, not all opinions received from the crowd are acceptable.

The constraints that we formally take into account can be broadly categorized into two types, namely main and background constraints.

\begin{itemize}
    \item \textbf{Main Constraints:}
The main constraints only involve the proposed sewage lines and deal with the area in which the sewage lines should be enclosed, minimum separability between two lines proposed by the same crowd worker, congestion among the finally proposed sewage lines and budget constraints.

\begin{enumerate}
    \item \textbf{Boundary:} If l is a proposed sewage line by a crowd worker then l should lie inside convex region $P$  
    \item \textbf{Intra-Separability:} If $l_1$ and $l_2$ are sewage lines suggested by the same crowd worker then any one of $l_1$ or $l_2$ should be rejected if $d_1(l_1,l_2) < D_1$ where $d_1$ denotes the constraint measure. This ensures that that each crowd worker gives $k^*$ many well-spaced opinions with no chances of repetition.
    \item \textbf{Intra-Congestion:} The final sewage lines proposed as the result of the consensus should be pairwise at least $D_2$ distance apart with respect to constraint measure.
    \item \textbf{Length:} To pertain to the Budget, the length of the final sewage lines proposed as the result of the consensus should be less than or equal to $L$. 
\end{enumerate}

\item \textbf{Background Constraints:}
The background constraints involve the existing sewage lines along with the proposed sewage lines and deal with connectivity, minimum separability of background sewage lines with proposed sewage lines and congestion of the new sewage lines to be constructed with existing ones.     

\begin{enumerate}
    \item \textbf{Intersection:} If $l$ is a suggested sewage line then $l$ should share at least one end point with one of the background sewage lines. This ensures that the sewage lines are well connected.
    \item \textbf{Inter-Separability:} If $l$ is a sewage line proposed by a crowd worker then $l$  should be rejected if $d_1(l,b) > D_1$, where $b$ is any existing sewage line. This ensures that the new sewage lines proposed maintains a suitable distance with the background sewage lines.
    \item \textbf{Inter-Congestion:} The final sewage lines proposed as the result of the consensus should be at least $D_2$ distance apart with respect to constraint measure, from every background sewage line. This ensures that the new sewage lines to be constructed are not congested with the existing sewage lines.
\end{enumerate}
\end{itemize}

It is important to understand that though we are considering the same distance measure for Separability and Congestion, they serve unique purposes. While Separability ensures that all opinion by the same crowd worker be distinguishable by a certain distance, Congestion constraint ensures the new sewage lines to be constructed are not too close to each other. Separability constraint is applied to remove erroneous data while Congestion constraint ensures proper sewage architecture.

\subsection{Data Collection}
The data was collected through a Google Form showing the following image that reflects the existing sewage lines (aka Cannels) of the drainage system in Greater Kolkata. Look at the coordinates on the map shown below and suggest coordinates (within the range (0, 0) to (42, 18)) for constructing new sewage lines. You should obey the following constraints.

\begin{enumerate}
    \item Your suggested sewage line should not be of length (Euclidean distance between the two end points) more than 10 units.
    \item Your suggested sewage lines should be at least 10 unit distant (minimum Euclidean distance) from each other.
    \item Your suggested sewage line should have an intersection with an existing Cannel. Provide a pair of coordinates (as (x1, y1), (x2, y2)) to denote the end points of the FIRST sewage line you suggest.
\end{enumerate}

\subsection{Handling the Background Constraints}
To handle the background constraints, we perform revisions of the input opinions.

\subsection{Model of Judgment Analysis}
We deal with the constraints at two phases of the proposed approach -- pre-processing and post-processing. While in the former phase, the proposed sewage lines are converted into valid opinions (to satisfy the constraints like Boundary, Intra-Separability, Inter-Separability and Intersection), in the latter one, the consensus opinions are corrected (to satisfy the constraints like Intra-Congestion, Inter-Congestion and Length). We propose an algorithm that segregates the entire approach into three parts, namely (i) filtering or adjusting the opinions, which violate the pre-Processing constraints. (ii) grouping the opinions yielding the centroids as the initial consensus of opinions, (iii) returning the final consensus of opinions that satisfy both pre-processing and post processing constraints.

We can simply visualize the step (ii) as finding the centroids of the opinions (lines), that satisfy the given constraints. So, the clustering of lines (and finding the respective medians) becomes the core of the approach. We use the k-medians inspired approach suggested by Marom et al. \cite{marom2019b}. The only exception is we use the \textbf{Hausdroff distance} in this case. The Hausdorff distance is a measure of similarity that calculates the greatest distance between any point in $X$ and the closest point in $Y$. In case we use the infimum distance as in \cite{marom2019b}, all sewage lines might get rejected as every sewage line need to have at least one connection with a background sewage line, and hence they will be rejected because of the inter-separability constraint. Therefore, we use the Hausdroff distance because it returns a 0 as the distance between a pair of lines iff the two lines are the same. The aforementioned approach is formally presented in Algorithm~1.

At the beginning, in \textbf{step 1}, the algorithm filters the sewage lines, which do not satisfy the boundary constraint. In \textbf{step 2}, the algorithm adjusts the sewage lines that do not satisfy the intersection constraint by extending from one end to the shortest line that at least one of its end points lie on some background sewage line. In \textbf{step 3}, the sewage lines that violate the constraint of Inter-Separability are filtered out. Subsequently, in \textbf{step 4}, if two sewage lines proposed by the same crowd worker say, $l_1$ and $l_2$ violate intra-separability, we remove one of them while retaining the other. In \textbf{step 5}, we run a centroid based clustering algorithm to get $k^*$ groups and medians as output that do not necessarily satisfy post-processing constraints. In \textbf{step 6}, we find the next best median in each group that satisfy the post-processing constraints of length, intra-congestion and  intra-congestion. The aforementioned algorithm is detailed below.

\begin{algorithm}\label{Algorithm}
\caption{Finding the consensus of sewage lines satisfying constraints}
\textbf{Input:} A convex region $P$, a number of $k$ background sewage lines $\mathcal{B} = \{O_1^0 = ((x_{11}^0, y_{11}^0), (x_{12}^0, y_{12}^0)), O_2^0 = ((x_{21}^0, y_{21})^0, (x_{22}^0, y_{22}^0)), \ldots, O_k^0 = ((x_{k1}^0, y_{k1}^0), (x_{k2}^0, y_{k2}^0))\}$ inside $P$, sets of sewage lines $\mathcal{S} = \{\{O_1^1, O_2^1, \ldots, O_{k^*}^1\}, \{O_1^2, O_2^2, \ldots, O_{k^*}^2\}, \ldots, \{O_1^n, O_2^n, \ldots, O_{k^*}^n\}\}$, suggested by $n$ crowd workers, respectively, where $O_i^j = ((x_{i1}^j, y_{i1}^j), (x_{i2}^j, y_{i2}^j))$.\\
\textbf{Output:} The consensus of the sewage lines satisfying pre-processing and post-processing constraints.\\
\textbf{Algorithmic Steps:}
\begin{algorithmic}[1]
\STATE $\mathcal{S}$ $\leftarrow$ \texttt{CONSTRAINT-BOUNDARY}($P$, $\mathcal{S}$)
\STATE $\mathcal{S}$ $\leftarrow$ \texttt{CONSTRAINT-INTERSECTION}($\mathcal{B}$, $\mathcal{S}$)
\STATE $\mathcal{S}$ $\leftarrow$ \texttt{CONSTRAINT-INTER-SEPARABILITY}($\mathcal{B}$, $\mathcal{S}$, $D_1$)
\STATE $\mathcal{S}$ $\leftarrow$ \texttt{CONSTRAINT-INTRA-SEPARABILITY}($\mathcal{S}$, $D_1$)
\STATE $\mathcal{G}$ $\leftarrow$ \texttt{CENTROID-BASED-CLUSTERING-OF-LINES}($\mathcal{S}$, $k^*$,$MAX\_ITER$)
\STATE $\mathcal{M}$ $\leftarrow$ \texttt{POST-PROCESSING OPTIMISATION}($\mathcal{G}, D_2, L$)
\RETURN $\mathcal{M}$
\end{algorithmic}
\end{algorithm}

\begin{enumerate}
\item \textbf{CONSTRAINT-BOUNDARY()}\\
This subroutine (see Technical Appendix) retains a sewage line $l$ if it lies within convex region $P$, otherwise rejects it. This is necessary so that the sewage lines outside the demarcated area should not contaminate the data.


\item\textbf{CONSTRAINT-INTERSECTION()}\\
This subroutine (see Technical Appendix) checks if at least one end point of every proposed sewage line lies on a background sewage line to ensure connectivity and such a line is extended from one end-point till it reaches the the nearest background sewage line. Another line is constructed by extending the other end point. The shortest among the two lines is retained. The algorithm slightly modifies the erroneous data points instead of filtering them which results in retaining a larger amount constraint satisfying data.


\item \textbf{CONSTRAINT-INTER-SEPARABILITY()}\\
This subroutine (see Technical Appendix) removes the proposed sewage lines that do not maintain sufficient gap with the existing sewage lines i.e if $d(l,b) < D_1$ for some $b \in \mathcal{B}$, $l$ is rejected. This ensures a sufficient gap is maintained with the background sewage lines.


\item  \textbf{CONSTRAINT-INTRA-SEPARABILITY()}\\
This subroutine (see Technical Appendix) checks if for any $l_1,l_2 \in \mathcal{S}$, $d_1(l_1, l_2) < D_1$, then the algorithm selects one of the two lines by giving post-processing constraints priority and executes one of the following three scenarios.

\begin{itemize}
    \item If exactly one of $l_1$ or $l_2$  is of greater than $L$, then discard it.

    \item Else let $n_1$ denote the number of sewage lines(proposed) that lie within $D_2$ distance of $l_1$ with respect to $d_1$. Similarly define $n_2$. If $n_2>n_1$, retain $l_1$. Else, if $n_1>n_2$ retain $l_2$. This basically removes the more congested line while retaining the less congested one.

    \item In all other circumstances, retain $l_1$
\end{itemize}

            

\item \textbf{CENTROID-BASED-CLUSTERING-OF-LINES()}\\
This subroutine (see Technical Appendix) segregates the proposed sewage lines into $k^*$ groups. Let us define the following {\em general distance measure} to be deployed as a part of a clustering technique (inspired from the k-medians algorithm) to group the sewage lines. If there are two line segments $l_1=(x_1,x_2)$ and $l_2=(y_1,y_2)$, where $x_1,x_2,y_1,y_2 \in \mathbb{R}^2$, let us define
\[
d_2(l_1,l_2) = 
\begin{cases}
  \min\{d(x_i,y_j), i, j = 1, 2\}, & \text{if $l_1$ does not} \\
     & \text{intersect $l_2$} \\
  0, & \text{otherwise}
\end{cases}
\]

The median of a group of line segments $l_1, l_2,...,l_n$ is  $l_i$, where $i=\{\j : \sum_{1 \leq k \leq n}d_2(l_k,l_j)$ is minimum over $ 1 \leq j \leq n \}$. Initially, $k^*$ many opinions are randomly selected as cluster centroids and each proposed line is assigned to the cluster of its nearest centroid. Then the medians of the clusters so formed are re-calculated, then the clusters are reassigned to the new cluster centres. This is repeated till the cluster centres no longer change or till the maximum number of iterations are over. The clusters are returned.


\item \textbf{POST-PROCESSING-OPTIMISATION()}\\
This subroutine (see Technical Appendix) optimises the choice of medians while abiding by the post-processing constraints.It takes $\mathcal{G}$(clusters), $\mathcal{M}$(medians) and $\mathcal{B}$(background sewage lines) as input and returns the final consensus. Cost function of a line segment $l \in$  cluster $C=\{l_1,l_2,...,l_n\}$ is $\sum_{1 \leq i \leq n}d_2(l,l_i)$. We make a list of line segments of length less than or equal to $L$ in each cluster and arrange them in increasing order of their cost function. For example in cluster $C_i$, let the list be $M_i=(l^{(i)}_1,l^{(i)}_2,...,l^{(i)}_{k_i})$, where $k_i$ denotes the number of lines of length less than or equal to $L$ and cost  function$(l^{(i)}_1) \leq $cost function$(l^{(i)}_2) \leq ... \leq$ cost function$(l^{(i)}_{k_i})$. The algorithm selects the final $k^*$  sewage lines (say, $m_1,m_2,...,m_{k^*}$) iteratively that satisfy both length and congestion constraints. The $m_i$s are selected as follows.

$m_1 = l^{1}_{i_1}$ where $i_1 = \min \{k: 1 \leq k \leq k_1$ and $\min\limits_{b \in \mathcal{B}} d_1(l_k,b) \geq D_2 \}$\\
$m_2 = l^{2}_{i_2}$ where $i_2 = \min \{k: 1 \leq k \leq k_2$ and $\min\limits_{b \in \mathcal{B} \bigcup \{m_1\}} d_1(l_k,b) \geq D_2 \}$\\
In general,\\
$m_j = l^{j}_{i_j}$ where $i_j = \min \{k: 1 \leq k \leq k_j$ and $\min\limits_{b \in \mathcal{B} \bigcup \{m_1, \ldots, m_{j-1}\}} d_1(l_k,b) \geq D_2 \}$ $\forall 1 \leq j \leq k^*$\\

Report $m_i$s as the final consensus of opinions.

We have assumed that such $m_i$s always exist because of the prior restriction that most of our data points abide by the post-processing constraint of length. However, if any such error arises, that would require the congestion parameter $D_2$ to be reduced and the $m_i$s re-valuated. 


\end{enumerate}

\subsection{Results and Observations}
We see that the consensus of the opinions (provided as preferred sewage lines to be constructed) is more realistically significant if background constraints are included. Moreover, it is also better to perform adjustment of constraints (wherever possible) instead of removal, thereby ensuring retention of more opinions that leads to a better consensus. As a whole, the inclusion of background and other constraints makes the consensus both practical and acceptable to the opinion providers. This was verified through follow up interviews and analyses.

\section{Conclusion}
In order to secure the collection of more responsible opinions, we investigate a novel dimension of constrained judgment analysis problem in the current work. It accommodates background restrictions in the judgment analysis problem based on the application scenario. Our experiments highlight that the background constraints are capable of making the annotators responsible in providing feasible and realistic opinions (not just random). The inclusion of background details while collecting the data also yield better results. It is interesting to note that given the sewage lines as input opinions (in case study 2), there is no significance of the ordering in which the inputs are provided. Therefore, the challenge of label correspondence that arises in making a consensus of points is not applicable here \cite{chatterjee2017smart}. Though we apply centroid-based techniques to obtain a consensus of opinions, other kind of clustering approaches might be extended to address this problem. One can also consider this problem to be extended to opinions taken in the form of polygons or other shapes in higher dimensional spaces.

\newpage

\section*{Technical Appendix}
\subsection*{Subroutines of the Algorithm used in Case Study 2}
The constraint handing and other procedures are formally provided below.

\begin{table}[htp]
\hrule
\hfill
\begin{flushleft}
\texttt{CONSTRAINT-BOUNDARY($P$, $\mathcal{S}$)}
\{
\begin{algorithmic}[1]
\FOR{$C \in \mathcal{S}$}
    \FOR{$c \in C$}
        \IF{$c$ lies outside $P$}
            \STATE Remove $c$ from $C$
        \ENDIF
       \ENDFOR
\ENDFOR 
\RETURN $\mathcal{S}$\\
\}
\end{algorithmic}
\end{flushleft}
\hrule
\label{CONSTRAINT-BOUNDARY}
\end{table}

\begin{table}[htp]
\hrule
\hfill
\begin{flushleft}
\texttt{CONSTRAINT-INTERSECTION($B$, $\mathcal{S}$)}
\{
\begin{algorithmic}[1]
\FOR{$C \in \mathcal{S}$}
    \FOR{$c \in C$}
        \IF{no end-points of $c$ lie on a background sewage line }
            \STATE Extend end-point 1 of $c$ till it touches some $b \in \mathcal{B}$. Call the new line $c_1$.Similarly define $c_2$
            \IF{$length(c_1) < length(c_2)$}
                \STATE Replace $c$ by $c_1$ in $C$ 
            \ELSE
                \STATE Replace $c$ by $c_2$ in $C$
             \ENDIF
        \ENDIF        
    \ENDFOR
\ENDFOR
\RETURN $\mathcal{S}$\\
\}
\end{algorithmic}
\end{flushleft}
\hrule
\label{CONSTRAINT-INTERSECTION}
\end{table}

\begin{table}[htp]
\hrule
\hfill
\begin{flushleft}
\texttt{CONSTRAINT-INTER-SEPARABILITY($B$, $\mathcal{S}$, $D_1$)}
\{
\begin{algorithmic}[1]
\FOR{$C \in \mathcal{S}$}
    \FOR{$c \in C$}
        \FOR{$b \in \mathcal{B}$}
            \IF{$d_1(c,b) < D$}
                \STATE remove $c$ from $C$
                \STATE break
            \ENDIF
        \ENDFOR
    \ENDFOR
\ENDFOR 
\RETURN $\mathcal{S}$\\
\}
\end{algorithmic}
\end{flushleft}
\hrule
\label{CONSTRAINT-INTER-SEPARABILITY}
\end{table}

\begin{table}[htp]
\hrule
\hfill
\begin{flushleft}
\texttt{CONSTRAINT-INRA-SEPARABILITY($\mathcal{S}, D_1$)}
\{
\begin{algorithmic}[1]
\FOR{$C \in \mathcal{S}$}
    \FOR{$c_1,c_2 \in C$}
        \IF{$d_1(c_1,c_2) < D_1$}
            \IF{$length(l_2) \leq L < length(l_1)$}
                \STATE remove $l_1$
            \ELSIF{$length(l_1) \leq L < length(l_2)$}
                \STATE remove $l_2$
            \ELSIF{$n_2>n_1$}
                \STATE remove $n_2$
            \ELSIF{$n_1>n_2$}
                \STATE remove $n_1$
            \ELSE
                \STATE remove $l_2$ 
            
            \ENDIF    
        \ENDIF
       \ENDFOR
\ENDFOR 
\RETURN $\mathcal{S}$\\
\}
\end{algorithmic}
\end{flushleft}
\hrule
\label{CONSTRAINT-INTRA-SEPARABILITY}
\end{table}

\begin{table}[h]
\hrule
\hfill
\begin{flushleft}
\texttt{CENTROID-BASED-CLUSTERING-OF-LINES($\mathcal{S}$, $MAX\_ITER$)}
\{
\begin{algorithmic}[1]
\STATE Make a list of all the proposed sewage lines and call it 
       $\mathcal{A}$
\STATE Randomly choose $k^*$ sewage lines $\{c_1, c_2, \ldots, c_{k^*}\} \in \mathcal{A}$
\STATE Let the clusters containing $\{c_1, c_2, \ldots, c_{k^*}\}$ be called $\{C_1, C_2, \ldots, C_{k^*}\}$ respectively.
\STATE $i$ $\leftarrow$ 0
\REPEAT
    \FOR{$a$ in $\mathcal{A}$}
        \STATE Assign $a$ to $C_i$ if $d_2(s,c_i)$ is minimum
    \ENDFOR
    \STATE Recompute $c_i$ as the median of the new $C_i's$
    \STATE $i$ $\leftarrow$ $i + 1$
\UNTIL{$c_i's$ no longer change or $i = MAX\_ITER$}
\STATE $\mathcal{G}=\{C_i, 1 \leq i \leq k^*\}$ and $\mathcal{M}=\{c_i, 1 \leq i \leq k^*\}$
\STATE return $\mathcal{G}$
\end{algorithmic}
\}
\end{flushleft}
\hrule
\label{CENTROID-BASED-CLUSTERING-OF-LINES}
\end{table}

\begin{table}[htp]
\hrule
\hfill
\begin{flushleft}
\texttt{POST-PROCESSING-OPTIMISATION($\mathcal{G},\mathcal{B},L, D_2$)}
\{
\begin{algorithmic}[1]
\FOR{$i \in \mathcal{N}_{k^*} $}
   \STATE Calculate $M_i$ for cluster $C_i$
   \FOR{$m \in M_i$}
       \IF{ $\min\limits_{b \in \mathcal{B} \bigcup \{m_1,...,m_{i-1}\}} d_1(m,b) \geq D_2$}
          \STATE $m_i \leftarrow m$
          \STATE \textbf{break}
       \ENDIF   
   \ENDFOR
\ENDFOR
\STATE $\mathcal{M} = (m_1,m_2,...,m_{k^*})$
\STATE return $\mathcal{M}$
\end{algorithmic}
\}
\end{flushleft}
\hrule
\label{POST-PROCESSING-OPTIMISATION}
\end{table}

\newpage

\subsection*{Complexity Analysis of the Algorithm used in Case Study 2}
\textbf{Time Complexity Analysis:} The time complexity begins with the \textbf{CONSTRAINT-BOUNDARY} component, which has a complexity of \( O(n \times k^*) \), where \( n \) represents the number of sets and \( k^* \) the line segments per set. The \textbf{CONSTRAINT-INTERSECTION} and \textbf{CONSTRAINT-INTER-SEPARABILITY} components both have complexities of \( O(n \times k^* \times k) \), reflecting the processing of each line against \( k \) background lines. A more intensive computation occurs in the \textbf{CONSTRAINT-INTRA-SEPARABILITY}, which checks pairwise distances within each set, resulting in \( O(n \times (k^*)^2) \). The most demanding part, however, is the \textbf{CENTROID-BASED-CLUSTERING-OF-LINES}, where the complexity rises to \( O(n \times (k^*)^2 \times \text{MAX ITER}) \), assuming multiple iterations for clustering. Lastly, the \textbf{POST-PROCESSING OPTIMISATION} introduces a complexity of \( O(n \times k^* \times \log(k^*)) \) for sorting operations. Consequently, the dominant time complexity in this analysis is from the centroid-based clustering, particularly due to the potential for numerous iterations, leading to an overall time complexity of \( O(n \times (k^*)^2 \times \text{MAX ITER}) \).

\textbf{Space Complexity Analysis:} Space complexities start minimal with \textbf{CONSTRAINT-BOUNDARY}, \textbf{CONSTRAINT-INTERSECTION}, \textbf{CONSTRAINT-INTER-SEPARABILITY}, and \textbf{CONSTRAINT-INTRA-SEPARABILITY}, each at \( O(1) \), operating in place without additional storage. However, the \textbf{CENTROID-BASED-CLUSTERING-OF-LINES} and \textbf{POST-PROCESSING OPTIMISATION} require more space, with complexities of \( O(n \times k^*) \) due to the need to store centroids and interim results. Thus, the overall space complexity, influenced mainly by these latter components, is \( O(n \times k^*) \).

\end{document}